\documentstyle[12pt]{article}
\begin{document}
\title{QUANTIZED FRACTAL SPACE TIME AND STOCHASTIC HOLISM}
\author{B.G. Sidharth$^*$\\
B.M. Birla Science Centre, Hyderabad 500 063 (India)}
\date{}
\maketitle
\footnotetext{$^*$E-mail:birlasc@hd1.vsnl.net.in}
\begin{abstract}
The space time that is used in relativistic Quantum Mechanics and Quantum
Field Theory is the Minkowski space time. Yet, as pointed out by several
scholars this classical space time is incompatible with the Heisenberg
Uncertainity Principle: We cannot go down to arbitrarily small space time
intervals, let alone space time points. Infact this classical space time is
at best an approximation, and this has been criticised by several scholars.
We investigate, what exactly this approximation entails. 
\end{abstract}
Over the past several decades, time has been studied by several scholars from
different perspectives\cite{r1}-\cite{r30}. As pointed out in Chapter 3, Newtonian space time
was purely geometrical, as compared to Einstein's physical study of it, whether
it be the Minkowski space time of Special Relativity or the Riemannian space time
of General Relativity. In almost all these studies, as also in Quantum Field
Theory, we still speak of space time points and deal with rigid scales even
though the Quantum Mechanical Uncertainty Principle contradicts these notions
as discussed in preceding Chapters like Chapters 2, 3 and 6.\\
In the preceding Chapters we have highlighted these shortcomings and have referred
to the concept of discrete space time. There is a nuance, though. Discrete
space time could still be thought of in the context of rigid minimum units, for
example the Planck length, or more generally the Compton wavelength of the
most massive elementary particle\cite{r31}.\\
However, as highlighted in Chapter 6 we have considered the Compton wavelength,
firstly without restricting it to the most massive particle, and secondly in
the context of a stochastic underpinning: We were lead to Quantized Fractal
Space Time (QFST).\\
Using QFST, we then saw in Chapter 6 that
\begin{equation}
(\vec x , t) \to (\vec x , t) \gamma\label{e1}
\end{equation}
where the $\gamma$'s are matrices. However the constraints imposed by the
commutation relations
$$[y, z] = (\imath a^2 / \hbar) L_{x,} [t, y] = (\imath a^2 / \hbar c)M_{y,}$$
$$[x, p_y] = [y, p_x] = \imath \hbar (a/ \hbar)^2 p_xp_y ;$$
require the $\gamma$'s to be atleast, $4 \times 4$ Dirac matrices. At the same time
these commutation relations underlie the double connectivity or spin half,
which is closely connected with non-commutativity, as we saw (Cf. also ref.\cite{r32}).
From equation (\ref{e1}) it would appear that there are extra dimensions - indeed
we encountered this in the previous Chapter. As pointed out, the situation is similar to that
in Quantum Superstrings. These extra dimensions are curled up or
supressed, in the unphysical Compton region. In the Kaluza-Klein theory, on
the other hand, the curling up takes place within the Planck length. Still,
we could reconcile the two.\\
We also saw in Chapter 6 that the double Weiner process within the Compton scale
gives rise to Special Relativity \underline{and} Quantum Mechanics, and indeed time
itself, through the equation
$$x \to x+\imath x' \quad \mbox{where}\quad x'=ct$$
and the complex velocity potential $V -\imath U$. (We also saw in
Chapter 3 that the so called non-relativistic Quantum Theory is not really
Galilean invariant, as a true non-relativistic theory should be. This is borne
out by the Sagnac effect \cite{r33}.)\\
On the other hand if there were no double Weiner process (or zitterbewegung),
then the diffusion constant $\nu$ would vanish and there would be neither
Special Relativity nor Quantum Mechanics (nor time!). This would also amount to the
disappearance of the quantized vortices in the hydrodynamical formulation seen
in Chapter 3.\\
These situations expose the mismatch of the classical ideas of space time and
Quantum Theory. Indeed as Wheeler put it \cite{r34}, "No prediction of space
time, therefore no meaning for space time is the verdict of the Quantum
Principle. That object which is central to all of Classical General Relativity,
the four dimensional space time geometry, simply does not exist, except in a
Classical approximation." Inspite of this, to understand the nature of the non-
Quantum Mechanical space time of Classical Theory, let us take the diffusion
constant of Chapter 6, $\nu$(or the Planck
constant $h$) to be very small, but non-vanishing. That is we consider the
semi classical case. This is because, a purely classical description, does not
provide any insight.\\
It is well known that in this situation we can use the WKB
approximation\cite{r35}. In this case the right hand side of the equation
$$\psi = \sqrt{\rho} e^{\imath /\hbar S}$$
goes over to, in the one dimensional case, for simplicity,
$$(p_x)^{-\frac{1}{2}} e^{\frac{\imath}{\hbar}\int p(x)dx}$$
so that we have, on comparison,
\begin{equation}
\rho = \frac{1}{p_x}\label{e2}
\end{equation}
In this case the condition $U = 0$ implies
\begin{equation}
\nu \cdot \nabla ln(\sqrt{\rho}) = 0\label{e3}
\end{equation}
(Cf. Chapter 6). This semi classical analysis suggests that
$\sqrt{\rho}$ is a slowly varying
function of $x$, infact each of the factors on the left side of (\ref{e3})
would be $\sim 0(h)$, so that the left side is $\sim 0(h^2)$ (which is
being neglected). Then from (\ref{e2}) we conclude that $p_x$
is independent of $x$, or is a slowly varying function of $x$.\\
The equation of continuity now gives
$$\frac{\partial \rho}{\partial t} + \vec \nabla (\rho \vec v ) =
\frac{\partial \rho}{\partial t} = 0$$
That is the probability density $\rho$ is independent or nearly so, not
only of $x$, but also of $t$. We are thus in a stationary and homogenous scenario.
This is strictly speaking, possible only in a single particle universe, or for a completely isolated
particle, without any effect of the environment. Under these circumstances we
have the various conservation laws and the time reversible theory, all this taken over
into Quantum Mechanics as well. With Wheeler's turn of phrase, though with
a slightly different connotation, this is "time without time" or "change
without change", an approximation valid for small, incremental changes, as
indeed is implicit in the concept of a differentiable space time manifold.
All this should not be surprising. In the
preceding Chapter, section 2, we have indeed noted the limitations of the
Field approach, a framework otherwise necessary for studying a multitude
of particles\cite{r36}.\\
We noted in Chapter 2, Prigogine's statement that "Our physical world is no
longer symbolised by the stable and periodic planetary motions that are at the
heart of classical mechanics".  The moment we consider the simplest of cases, viz., the three body problem, even in
Newtonian Mechanics, it amounts to bringing in instabilities due to the environmental or holistic
feature.\\
In our considerations in the preceding Chapters, we encountered exactly this
holistic feature in a stochastic setting, what may be called stochastic
holism. For example this is embodied not only in the various Large Number
relations deduced in Chapter 7 or the fluctuations underlying interactions in Chapter 8, but also in the fact that as we saw in the last
Chapter, the number of particles in the universe (and a maximal universal
velocity) can be considered to be the only free parameter - the other
microphysical constants are dependent on this. If Classical Theory can be
compared to a strucutre constructed with rigid building blocks, with ideas
like local realism thrown in\cite{r37}, we are here
talking about a picture where the building blocks themselves depend on the
overall structure stochastically.\\
In this case the puzzle of the irreversibility of time, as discussed by several
scholars\cite{r38}-\cite{r46} also disappears. Irreversibility is a consequence of the
statistical, or in a manner of speaking Thermodynamic nature of space time.
Indeed we saw in Chapter 6 and Chapter 7 that the equation
$$T = \sqrt{N}\tau$$
provides an immediate arrow of time, while in Chapter 10 we saw how QFST could
explain the Kaon decay.\\
It was shown in Chapter 7 that this picture of fluctuations in the context of
QFST leads automatically to the Large Number Relations. Ofcourse it is possible
to trivialise these Large Number relations as coincidences with an anthropic type
argument \cite{r47}. Our approach on the other hand has been not so much an
explanation for cosmic numerology, as it has been a search for minimum
underlying simple principles, in the spirit of Occam's razor, that would
explain disparate phenomena. That is what science is all about.

\end{document}